\documentstyle[12pt]{article}
\newcommand{\bb}{\begin{eqnarray}}
\newcommand{\ee}{\end{eqnarray}}
\begin{document}
\title{{Entropy of near-extremal dyonic black holes}}
\author{P. Mitra\thanks{e-mail mitra@tnp.saha.ernet.in}\\
Saha Institute of Nuclear Physics\\
Block AF, Bidhannagar\\
Calcutta 700 064, INDIA}
\date{hep-th/9712252}
\maketitle
\begin{abstract}
In this note it is shown that near-extremal four dimensional dyonic black
holes, where the dilaton is not constant, can be described by a microscopic
model consisting of a one-dimensional gas of massless particles.
\end{abstract}
\bigskip
\newpage
\section{Introduction}
Traditional black hole thermodynamics taught us that a black hole
has an entropy given by a quarter of the area of the horizon, but
the statistical mechanics of this entropy remained something of a
puzzle. This is because of  the  uncertainties  involved  in  the
quantization  of  gravity.  The semiclassical interpretations of
the area law were not  microscopic enough. Recently, however, in  the
context  of  the string theory approach to gravity, a description
of black hole states in terms of D-branes has been found to allow
a  derivation  of  the area law from a counting of quantum states
\cite{sv}.

Since then, there have  been  many  studies  on  the  microscopic
description  of  black holes. Many  of  these  studies  involved extremal
and near-extremal Reissner -  Nordstr\"{o}m black holes in different
dimensions. The simpler Schwarzschild  black  hole  could  not  be
accommodated in  those studies essentially because it has no extremal
version.  The so-called dilatonic black hole  may  come  to  mind  in  this
connection,  but though there are extremal dilatonic black holes, these
have  zero  horizon  area.  A  satisfactory   microscopic interpretation  of
such  black  holes has not been possible. The next in the order of
increasing  complication  is  probably  a dyonic black hole with both
electric and magnetic charges.  Dyonic  string-based black holes have often
been considered in the literature.  Those  with  at  least two magnetic and
two electric charges have horizons of non-zero area. In fact  the  Reissner
-  Nordstr\"{o}m black  hole  is  a special case of this class of black
holes with the two magnetic and two electric charges all equal.

The explicit dyonic solutions of the low energy field equations  arising
from toroidally compactified heterotic string theory were given for the
extremal case in \cite{CY1} and for the non-extremal case in \cite{CY2}. In
\cite{LW} it was argued that the number of microscopic (string) states
corresponding to a macroscopic state of the extremal black hole would give
rise to an entropy equal to a quarter of the area of the horizon, that is,
what one  could  expect from  traditional black hole thermodynamics.
However, as pointed out in \cite{GM1}, the detailed argument in \cite{LW}
involved a clustering hypothesis which is known to fail even for the simpler
black holes which are only electrically charged. This made the claim quite
uncertain.  It was further pointed out in \cite{GM1} that the semiclassical
value of the entropy for only a non-extremal black hole is a quarter of the
area, that is normally not  the case for an extremal black hole. In  fact,
if  the  quantum  extremal  black hole is defined by starting with a
classical extremal black hole and then quantizing it, the  (semiclassical)
entropy  appears  to  be  of degree  one in  the charges, unlike the area,
which is of degree two.

Meanwhile, improved techniques of counting D-brane states indicated that the
extremal dyonic black hole indeed can be assigned a microscopic entropy
equal to a quarter of the area \cite{MS1}. This would seem to be in
disagreement with the semiclassical result \cite{GM1}, but it was soon
realized that the order of the processes of quantization and extremalization
is important \cite{GM2}, so that if instead of quantizing an extremal,
classical black hole, the extremality was imposed {\it after} functional
integration over both extremal and non-extremal topologies, a different
result could be obtained. In the case of Reissner - Nordstr\"{o}m black
holes, the area formula was explicitly derived in this way \cite{GM2}.  This
indicated  that  the microscopic area formula  could  be accommodated with
semiclassical ideas even for extremal black holes.

The interest then shifts to non-extremal dyonic black holes. These black
holes are  dilatonic,  and  do  not  fall  into the category \cite{KT} of
black hole solutions with constant dilaton, which allow reliable microscopic
state counting away from extremality.  Although the dilaton is not in
general constant in these black holes, it does not diverge near the horizon,
so  it  is  not entirely  unreasonable to hope for a microscopic calculation
to make sense even in  the  near-extremal  case. Indeed, such a calculation
has been given for some special ranges of values of the charges \cite{HLM}.
The purpose of the present note is  to confirm that near-extremal dyonic
black holes with {\it generic} values of the charges can be described by
microscopic models.

We shall not make explicit use of any string model to fit the thermodynamics
of near-extremal dyonic black holes.  As pointed out in \cite{alwis}, a gas
of massless particles in one space dimension can be used for this purpose.
This is essentially the same as using a conformal field theory, as suggested
in \cite{MS2}. A detailed formulation of the agreement between the gas model
and the Reissner - Nordstr\"{o}m black hole was presented in \cite{puri},
and it will be extended here to the case  of  dyonic  black  holes  with
generic values of the charges.

In section 2 the reader is reminded about dyonic black holes \cite{CY1,
CY2}.  In section 3 the one-dimensional gas model is applied to these black
holes. In the concluding discussion some remarks are made about the
difference of the formulations given here and in \cite{alwis}.

\section{Dyonic black holes}
A simple four-dimensional dyonic black hole solution of toroidally
compactified heterotic string theory can be described by the
metric \cite{CY2}
\bb
ds^2=-\lambda dt^2 + {dr^2\over\lambda} + R^2 d\Omega^2,
\ee
where
\bb
\lambda &=&{(r+b)(r-b)\over R^2},\nonumber\\
R^4&=&(r+\hat Q_1)(r+\hat Q_2)(r+\hat P_1)(r+\hat P_2),\nonumber\\
\hat Q_1&=&\sqrt{Q_1^2+b^2}, ~etc.
\label{met}\ee
It carries two independent electric charges $Q_1,Q_2$ with the same sign
as well as  two independent magnetic charges
$P_1,P_2$ with the same sign, 
and there is a real positive parameter $b$ which specifies the position
of the horizon and measures the departure from extremality. The expressions
for the 28 gauge fields to which the 28-component charges (with not all
components independent in this solution) are coupled will be omitted
for simplicity. The expression for the dilaton field is
\bb
e^{2\phi}={(r+\hat P_1)(r+\hat P_2)\over (r+\hat Q_1)(r+\hat Q_2)},
\ee
which shows that the dilaton does not in general vanish for these solutions,
and is not even constant, but goes to zero asymptotically as $r\to\infty$.

The horizon of such a black hole has a finite area given by
\bb
A=4\pi\sqrt{
(b+\hat P_1)(b+\hat P_2)(b+\hat Q_1)(b+\hat Q_2)},
\ee
and it remains nonvanishing in the extremal limit provided all four
charges are nonvanishing. The Hawking temperature is
\bb
T_H={2b\over A}.
\ee
It vanishes in the extremal limit. The ADM mass of the black hole is
\bb
M&=&{1\over 4}(\hat Q_1+\hat Q_2+\hat P_1+\hat P_2)\nonumber\\&\ge & M_{ex}
={1\over 4}( |Q_1+ Q_2|+ |P_1+ P_2|).
\ee

\section{Microscopic model for near-extremal black holes}
In this section we seek to use the one-dimensional
gas of massless particles (cf. \cite{alwis, puri})
as a model for the dyonic black  holes. The model can work for black
holes in  any  number  of
dimensions, though it will be used here only for four dimensional black holes.
The   particles   can   be  either  left-moving  or
right-moving -- there is no mixing between the two types. Both bosons
and fermions  can  be  present.  If  the  total  length  of  the
one-dimensional  space is $L$, the entropy and the energy are given
by
\bb
S&=&{\pi L\over 6\hbar}[n_LT_L+n_RT_R],\nonumber~\\
E&=&{\pi L\over 12\hbar}[n_LT_L^2+n_RT_R^2],
\ee
where $n_L(n_R)$ is the number of left(right)-moving bosons  plus
half  the  corresponding  number  of  fermions, because a fermion contributes
only a half of the contribution a boson makes to the free energy.
In the absence of
interactions,  the  left  and  right  degrees  of  freedom   are
independent, and the corresponding temperatures can be different.
The  effective  temperature  may  be   defined   by   $({\partial
S\over\partial E})^{-1}$, the differentiation being carried out at
constant momentum, {\it i.e.,} constant difference between $E_L$
and $E_R$. This leads to a temperature
\bb
T={2T_LT_R\over T_L+T_R}
\ee
equal to the harmonic mean of $T_L$ and $T_R$.
If $n_L=n_R=n$, these equations get somewhat simplified and by
eliminating $T_L, T_R$, one gets the single relation
\bb
E={\pi nL\over 12\hbar}\bigg[\bigg({6\hbar S\over \pi nL}\bigg)^2-
{6\hbar ST\over \pi nL}\bigg].\label{E}
\ee

To compare these quantities with those for a near-extremal dyonic black hole
in four dimensions, it is necessary to fix two parameters of the gas to be
equal to the corresponding parameters of the black hole, and check how
the third parameter of the gas compares with that parameter for the black
hole. Thus, we could set the temperatures and energies to be equal
and compare  the entropies. However, as the entropy of the gas
enters the energy
quadratically, the expression for the entropy in terms of the energy
involves a square root. It can be handled, but it is much
simpler to match the temperatures and the entropies and compare the
energies. Thus we put
\bb
T&=&T_H(b,Q_1,Q_2,P_1,P_2)\nonumber\\
 &=&{b\over 2\pi\sqrt{\prod Q}}\bigg(1-{b\over 2}\sum{1\over |Q|} 
+ {b^2\over
8}\bigg(\sum{1\over |Q|}\bigg)^2\bigg)+\cdots,\nonumber \\
S&=&{A(b,Q_1,Q_2,P_1,P_2)\over 4}\nonumber\\
  &=&\pi\sqrt{\prod Q}\bigg( 1+{b\over 2}\sum{1\over |Q|} + {b^2\over
8}\bigg(\sum{1\over |Q|}\bigg)^2\bigg)+\cdots
\ee
in (\ref{E}) to get
\bb
E={3\pi\prod Q\over nL}+b({3\pi\prod Q\sum {1\over |Q|}\over nL}-{1
\over 4}) + b^2\cdot{3\pi\prod Q(\sum{1\over |Q|})^2\over 2nL}+\cdots.
\ee
Here sums and products over $Q$ are understood to include both the
electric and both the magnetic charges and $b$ is understood to be small. 
This expression for the energy of the gas shows a quadratic increase with
$b$ (if higher degree terms are neglected), which is similar to the behaviour
of the blackhole mass 
\bb
4M=\sum |Q|+{b^2\over 2}\sum{1\over |Q|}+\cdots,
\ee
but it involves the characteristic $nL$ of the model and is not 
manifestly identical to
$M$. However, for consistency, we must require that the temperature
${dE\over dS}$ vanish for $b=0$, which is the point of extremality,
at which the temperature $T_H$ vanishes.
This yields the relation
\bb
{3\pi\prod Q\sum {1\over |Q|}\over nL}={1\over 4},\label{L}
\ee
between the product $nL$ and the
parameters $Q_1,Q_2,P_1,P_2$ characterizing the family of
black holes being considered.
$E$ and $M$ then agree in both the $b$ and the $b^2$ terms,
{\it i.e.,} for all $b$ for which $b^3$
can be neglected:
\bb
E=M+{1\over 4}({1\over\sum{1\over |Q|}}-\sum |Q|)+O(b^3).
\ee
The difference between $E$ and $M$ represents a zero-point shift depending
only on the charges of the black hole and independent of $b$.  The
gas of massless particles then can  be  regarded  as  a  model,   as far
as thermodynamics is concerned, for   the {\it family} of near-extremal
dyonic black   holes with a fixed set of values of the charges but varying
$b$. 

\section{Conclusion}
A microscopic model has thus been constructed for near-extremal dyonic black
holes in terms of the one dimensional gas suggested in \cite{alwis}. It was
of course envisaged in \cite{alwis} that the model would work for a general
class of black holes, but it was noted there that the energy of  the  gas
and the mass of the general black hole with the same temperature and entropy
could be made to agree up to a zero point energy, which is apparently {\it
dependent on the non-extremality parameter}, and thus not constant for the
family of near-extremal black holes.  The zero point shift  in the present
formulation is independent of the non-extremality parameter, so that the
{\it energy differences match}. This is what was demonstrated for the
Reissner - Nordstr\"{o}m case in \cite{puri} and the same behaviour has been
shown  here  for  generic  dyonic black holes. The new result is to be
regarded as a generalization of  the  earlier one because, as pointed out
above, the Reissner - Nordstr\"{o}m black hole is a special case  of  a
dyonic  black hole.   Applications  to  the  Schwarzschild  black  hole  or
the dilatonic one, which are also special  cases,  are  not  possible
because  in  the  former case there is no extremal limit, and in the latter
case the limit is singular.

It is instructive to compare the formulation of \cite{alwis} with
that  given here and understand why the zero point energies seem
to  behave  differently  even though  the  dyonic  black  hole  is
included in the class of black holes considered in \cite{alwis}.
That class is characterized, in four dimensions, for the case of
four charges, by the metric
\bb
ds^2=-\tilde\lambda dt^2 + {d\tilde r^2\over\tilde\lambda} +
\tilde R^2 d\Omega^2,
\ee
with
\bb
\tilde\lambda &=&{\tilde r(\tilde r-r_0)\over\tilde R^2},\nonumber\\
\tilde R^4&=&(\tilde r+ r_1)(\tilde r+ r_2)(\tilde r+ r_3)(\tilde
r+ r_4).
\ee
To  put  the  dyonic  metric  (\ref{met})  in  this  form,
one matches
\bb
r^2-b^2=\tilde r(\tilde r-r_0), ~ r+\hat Q_1=\tilde r+r_1, ~etc.
\ee
These imply that
\bb
r+b=\tilde r, ~ 2b=r_0, ~ \hat Q_1-b=r_1, ~etc.
\ee
The first of these simply means that  $\tilde  r$  is  translated
with respect to $r$, but the last one indicates that $r_1$ is not
just the charge $Q_1$ but  $\sqrt{Q_1^2+b^2}-b$,  which  involves
both  the  charge $Q_1$ and the non-extremality parameter $b$. It
is because of this that the dependence of the zero  point  energy
on the non-extremality parameter seems to be different in the two
formulations. There  is  in  fact  no
difference,   and one can identify the family of
near-extremal black holes with fixed charges and the thermal states  of
the gas.

\section*{Acknowledgment}
It is a pleasure to  acknowledge discussions with Amit Ghosh
on the gas model.

\end{document}